\documentclass[12pt]{article}
\usepackage{epsfig}
\usepackage{wrapfig}
\usepackage{subfigure}
\usepackage{amsmath}
\usepackage{hhline}
\usepackage{amssymb}
\usepackage{times}
\usepackage{cite}

\usepackage[]{lineno}

\newcommand{\be}{\begin{equation}}
\newcommand{\ee}{\end{equation}}
\newcommand{\la}{\langle}
\newcommand{\ra}{\rangle}

\newlength{\dinwidth}
\newlength{\dinmargin}
\begin{document}  
\newcommand{\pom}{{I\!\!P}}
\newcommand{\reg}{{I\!\!R}}
\newcommand{\slowpi}{\pi_{\mathit{slow}}}
\newcommand{\fiidiii}{F_2^{D(3)}}
\newcommand{\fiidiiiarg}{\fiidiii\,(\beta,\,Q^2,\,x)}
\newcommand{\n}{1.19\pm 0.06 (stat.) \pm0.07 (syst.)}
\newcommand{\nz}{1.30\pm 0.08 (stat.)^{+0.08}_{-0.14} (syst.)}
\newcommand{\fiidiiiful}{F_2^{D(4)}\,(\beta,\,Q^2,\,x,\,t)}
\newcommand{\fiipom}{\tilde F_2^D}
\newcommand{\ALPHA}{1.10\pm0.03 (stat.) \pm0.04 (syst.)}
\newcommand{\ALPHAZ}{1.15\pm0.04 (stat.)^{+0.04}_{-0.07} (syst.)}
\newcommand{\fiipomarg}{\fiipom\,(\beta,\,Q^2)}
\newcommand{\pomflux}{f_{\pom / p}}
\newcommand{\nxpom}{1.19\pm 0.06 (stat.) \pm0.07 (syst.)}
\newcommand {\gapprox}
   {\raisebox{-0.7ex}{$\stackrel {\textstyle>}{\sim}$}}
\newcommand {\lapprox}
   {\raisebox{-0.7ex}{$\stackrel {\textstyle<}{\sim}$}}
\def\gsim{\,\lower.25ex\hbox{$\scriptstyle\sim$}\kern-1.30ex%
\raise 0.55ex\hbox{$\scriptstyle >$}\,}
\def\lsim{\,\lower.25ex\hbox{$\scriptstyle\sim$}\kern-1.30ex%
\raise 0.55ex\hbox{$\scriptstyle <$}\,}
\newcommand{\pomfluxarg}{f_{\pom / p}\,(x_\pom)}
\newcommand{\dsf}{\mbox{$F_2^{D(3)}$}}
\newcommand{\dsfva}{\mbox{$F_2^{D(3)}(\beta,Q^2,x_{I\!\!P})$}}
\newcommand{\dsfvb}{\mbox{$F_2^{D(3)}(\beta,Q^2,x)$}}
\newcommand{\dsfpom}{$F_2^{I\!\!P}$}
\newcommand{\gap}{\stackrel{>}{\sim}}
\newcommand{\lap}{\stackrel{<}{\sim}}
\newcommand{\fem}{$F_2^{em}$}
\newcommand{\tsnmp}{$\tilde{\sigma}_{NC}(e^{\mp})$}
\newcommand{\tsnm}{$\tilde{\sigma}_{NC}(e^-)$}
\newcommand{\tsnp}{$\tilde{\sigma}_{NC}(e^+)$}
\newcommand{\st}{$\star$}
\newcommand{\sst}{$\star \star$}
\newcommand{\ssst}{$\star \star \star$}
\newcommand{\sssst}{$\star \star \star \star$}
\newcommand{\tw}{\theta_W}
\newcommand{\sw}{\sin{\theta_W}}
\newcommand{\cw}{\cos{\theta_W}}
\newcommand{\sww}{\sin^2{\theta_W}}
\newcommand{\cww}{\cos^2{\theta_W}}
\newcommand{\trm}{m_{\perp}}
\newcommand{\trp}{p_{\perp}}
\newcommand{\trmm}{m_{\perp}^2}
\newcommand{\trpp}{p_{\perp}^2}
\newcommand{\alp}{\alpha_s}
\newcommand{\etamax}{\eta_{\rm max}}

\newcommand{\alps}{\alpha_s}
\newcommand{\sqrts}{$\sqrt{s}$}
\newcommand{\LO}{$O(\alpha_s^0)$}
\newcommand{\Oa}{$O(\alpha_s)$}
\newcommand{\Oaa}{$O(\alpha_s^2)$}
\newcommand{\PT}{p_{\perp}}
\newcommand{\JPSI}{J/\psi}
\newcommand{\sh}{\hat{s}}
\newcommand{\uh}{\hat{u}}
\newcommand{\MP}{m_{J/\psi}}
\newcommand{\PO}{I\!\!P}
\newcommand{\xbj}{x}
\newcommand{\xpom}{x_{\PO}}
\newcommand{\ttbs}{\char'134}
\newcommand{\xpomlo}{3\times10^{-4}}  
\newcommand{\xpomup}{0.05}  
\newcommand{\dgr}{^\circ}
\newcommand{\pbarnt}{\,\mbox{{\rm pb$^{-1}$}}}
\newcommand{\gev}{\,\mbox{GeV}}
\newcommand{\WBoson}{\mbox{$W$}}
\newcommand{\fbarn}{\,\mbox{{\rm fb}}}
\newcommand{\fbarnt}{\,\mbox{{\rm fb$^{-1}$}}}
\newcommand{\dsdx}[1]{$d\sigma\!/\!d #1\,$}
\newcommand{\eV}{\mbox{e\hspace{-0.08em}V}}
%
%
\newcommand{\qsq}{\ensuremath{Q^2} }
\newcommand{\gevsq}{\ensuremath{\mathrm{GeV}^2} }
\newcommand{\et}{\ensuremath{E_t^*} }
\newcommand{\rap}{\ensuremath{\eta^*} }
\newcommand{\gp}{\ensuremath{\gamma^*}p }
\newcommand{\dsiget}{\ensuremath{{\rm d}\sigma_{ep}/{\rm d}E_t^*} }
\newcommand{\dsigrap}{\ensuremath{{\rm d}\sigma_{ep}/{\rm d}\eta^*} }

\newcommand{\dstar}{\ensuremath{D^*}}
\newcommand{\dstarp}{\ensuremath{D^{*+}}}
\newcommand{\dstarm}{\ensuremath{D^{*-}}}
\newcommand{\dstarpm}{\ensuremath{D^{*\pm}}}
\newcommand{\zDs}{\ensuremath{z(\dstar )}}
\newcommand{\Wgp}{\ensuremath{W_{\gamma p}}}
\newcommand{\ptds}{\ensuremath{p_t(\dstar )}}
\newcommand{\etads}{\ensuremath{\eta(\dstar )}}
\newcommand{\ptj}{\ensuremath{p_t(\mbox{jet})}}
\newcommand{\ptjn}[1]{\ensuremath{p_t(\mbox{jet$_{#1}$})}}
\newcommand{\etaj}{\ensuremath{\eta(\mbox{jet})}}
\newcommand{\detadsj}{\ensuremath{\eta(\dstar )\, \mbox{-}\, \etaj}}

\newcommand{\PR}{Phys. Rev.\ }
\newcommand{\PRE}{Phys. Rep. \ }
\newcommand{\PRL}{Phys. Rev. Lett.\ }
\newcommand{\PL}{Phys. Lett.\ }
\newcommand{\NP}{Nucl. Phys.\ }
\newcommand{\ZP}{Z. Phys.\ }
\newcommand{\AP}{Ann. Phys.}
\newcommand{\Coll}{Collaboration}
\newcommand{\CPC}{Comp. Phys. Comm. \ }

\newcommand{\clDq}{{\cal D}_q}

\def\Journal#1#2#3#4{{#1} {\bf #2} (#3) #4}
\def\NCA{\em Nuovo Cimento}
\def\NIM{\em Nucl. Instrum. Methods}
\def\NIMA{{\em Nucl. Instrum. Methods} {\bf A}}
\def\NPB{{\em Nucl. Phys.}   {\bf B}}
\def\PLB{{\em Phys. Lett.}   {\bf B}}
\def\PRL{\em Phys. Rev. Lett.}
\def\PRD{{\em Phys. Rev.}    {\bf D}}
\def\ZPC{{\em Z. Phys.}      {\bf C}}
\def\EJC{{\em Eur. Phys. J.} {\bf C}}
\def\CPC{\em Comp. Phys. Commun.}

\begin{titlepage}

\noindent

\noindent
\noindent

\vspace{2cm}
\begin{center}
\begin{Large}

{\bf Factorial Moments in Complex Systems
}
\end{Large}

\vspace{2cm}

Laurent Schoeffel \\~\\
CEA Saclay, Irfu/SPP, 91191 Gif/Yvette Cedex, \\
France
\end{center}

\vspace{2cm}

\begin{abstract}

Factorial
moments
are convenient tools  in particle physics to characterize the multiplicity
distributions when phase-space resolution ($\Delta$) becomes small.
They include all correlations within the system of
particles and represent integral characteristics of any correlation
between these particles.
In this letter, we show a direct comparison between high energy
physics  and quantitative finance results.
Both for physics and finance, we illustrate that correlations
between particles lead to a broadening of the multiplicity distribution 
and to dynamical fluctuations when the resolution becomes small enough.
From the generating function of factorial moments, we 
make a prediction on the gap probability  for sequences of returns
of positive or negative signs.
The gap is defined as the number of consecutive positive returns after a negative return,
thus this is a gap in negative return. Inversely for a gap in positive return.
Then, the gap
probability is shown to be exponentially suppressed within the gap size.
We confirm this prediction with data.

\end{abstract}

\vspace{1.5cm}

\begin{center}
\end{center}

\end{titlepage}

%
%
%
%

\section{Introduction}

Multiplicity distributions and correlations between
final-state particles  in nuclear interactions
are an important testing ground for  
analytic perturbative theory,  
as well as for Monte-Carlo (MC)  models  
describing the hadronic final state \cite{rew}.  
Two-particle 
angular correlations have been extensively studied experimentally
\cite{angul}. 
Specific statistical tools, namely the 
normalized factorial moments,
 have emerged in order to analyze in much details multiplicity distributions
measured in restricted phase-space regions.
The 
normalized factorial moments are defined as
$$
F_q(\Delta)=\la n(n-1)\ldots (n-q+1)\ra / \la n \ra^q,
\qquad q=2,3, \ldots ,  
$$
for a specified phase-space region of size $\Delta$. 
The number, $n$, of particles is measured inside  $\Delta$
and angled brackets $\la\ldots\ra$ denote 
the average over all events. The factorial
moments, along with cumulants \cite{cum} and bunching parameters \cite{bp},
are convenient tools  to characterize the multiplicity
distributions when $\Delta$ becomes small. Indeed,
for uncorrelated particle production within $\Delta$, 
Poisson  statistics
holds  and $F_q=1$ for all $q$. Correlations
between particles lead to a broadening of the multiplicity distribution 
and to dynamical fluctuations. In this case, the normalized factorial 
moments increase  
with decreasing $\Delta$. 
This  effect is frequently called intermittency \cite{bias}.  
As a matter of fact, it has  
been noticed in \cite{bias} that the use of  
factorial moments allows to extract the dynamical signal from the  
Poisson noise in the analysis of the multiplicity  
signal in high energy reactions. 

In addition, it has been shown that it is possible to define and compute a multi-fractal  
dimension, $\clDq$, for the theory of strong interactions \cite{DD93,OW93}
\begin{equation}
\label{inter}
F_q(\Delta)=\la n(n-1)\ldots (n-q+1)\ra / \la n \ra^q \propto \Delta^{(q-1)(1-{\cal D}_q/d)} 
\end{equation}
where d is the dimension of the phase space under consideration  
($d=2$ for the
whole angular phase space, and $d=1$ if one has integrated over, say  
the
azimuthal angle).   
In the constant coupling case $\clDq$ is well defined and
reads
\begin{equation}
\label{diminter}
{\cal D}_q = \gamma_0\frac{q+1}{q} 
\end{equation}
where $\gamma_0^2=4C_A{\alpha_S/ 2\pi}$, $\alpha_S$ is the strong interaction  
coupling constant, $C_A$ is the gluon color factor
\cite{DD93,OW93}.   
The choice of the factorial moments as a specific tool in order to study  
the scaling behavior of the high energy multiplicity  
distributions is then useful to analyze the underlying dynamics of the
processes. In principle, we can  extend this last idea to other fields where
factorial moments can be defined.

\section{Generating Function for Factorial Moments}
\label{formalism}

The multiplicity distribution is defined as
$
  P_{n} = {\sigma_{n}}/{\sum_{n=0}^{\infty}\sigma_{n}}    
$,
where $\sigma_{n}$ is the cross section of an
$n$-particle production process 
(the so-called topological cross section)
and the sum is over all possible values of $n$ so that
\begin{equation}
  \sum_{n=0}^{\infty}P_{n} = 1 .          \label{2}
\end{equation}
The  generating function can  be
defined as
\begin{equation}
  G(z) = \sum_{n=0}^{\infty }P_{n}(1+z)^{n}  ,    \label{3}
\end{equation}
which substitutes an analytic function of $z$ in place of the set of numbers 
$P_{n}$.
Then, we obtain the factorial moment or order $q$ as
\begin{equation}
  F_{q} = 
\frac {1}{\langle n \rangle ^{q}}\cdot \frac {d^{q}G(z)}{dz^{q}}\vline _{z=0}, 
\label{4}
\end{equation}
and the corresponding definition for cumulants
\begin{equation}
  K_{q} = \frac {1}{\langle n \rangle ^{q}}\cdot \frac {d^{q}\ln G(z)}{dz^{q}}.
\vline _{z=0}, \label{5}
\end{equation}
 
The expression for $G(z)$ can then be re-written as
\begin{equation}
  G(z) = \sum _{q=0}^{\infty } \frac {z^q}{q!} \langle n \rangle ^{q} F_{q} 
  \;\;\;\; ( F_0 = F_1 = 1 ),    \label{7}
\end{equation}
\begin{equation}
  \ln G(z) = \sum _{q=1}^{\infty } \frac {z^q}{q!} \langle n \rangle ^{q} K_{q} 
  \;\;\;\; ( K_1 = 1 ).    \label{8}
\end{equation}
The physical meaning of these moments has been discussed in
the previous section. Another interpretation can be seen 
from the above definitions if they are presented in the form of 
integrals of correlation functions. 
Let the single symbol $y$ represent all kinematic variables needed 
to specify the position of each particle in
the phase space volume~$\Delta $ \cite{d1}. 
A sequence of inclusive 
$q$-particle differential cross sections $d^{q}\sigma /dy_{1}\ldots dy_{q}$
defines the factorial moments as
\begin{equation}
  F_{q} \sim \frac{1}{\langle n\rangle ^q}\int _{\Delta }dy_{1} 
  \ldots \int _{\Delta } dy_{q}\frac {d^{q}\sigma }{dy_{1}\ldots dy_{q}}. 
\label{12b}
\end{equation}
Therefore, factorial
moments include all correlations within the system of
particles under consideration. 
They represent integral characteristics of any correlation
between the particles whereas cumulants of rank $q$ represent genuine
$q$-particle correlations not reducible to the product of lower order
correlations.

\section{Experimental Applications}

In Ref. \cite{lolo}, it has been shown that factorial moments can be applied conveniently
to quantitative finance.
Namely,
if we divide price series $y(t)$ in consecutive time windows of lengths $\Delta$,
we can define a set of events.
In each window, we have a certain number of positive returns  $n_+$,
where $y(t)-y(t-1) >0$, and similarly of
negative returns $n_-$.
If the sequence of returns is purely
 uncorrelated, following a  
Gaussian  statistics at all scales,
we expect $F_q=1$ for all $q$. 

\begin{figure}[htbp]
  \begin{center}
    \includegraphics[width=0.4\textwidth]{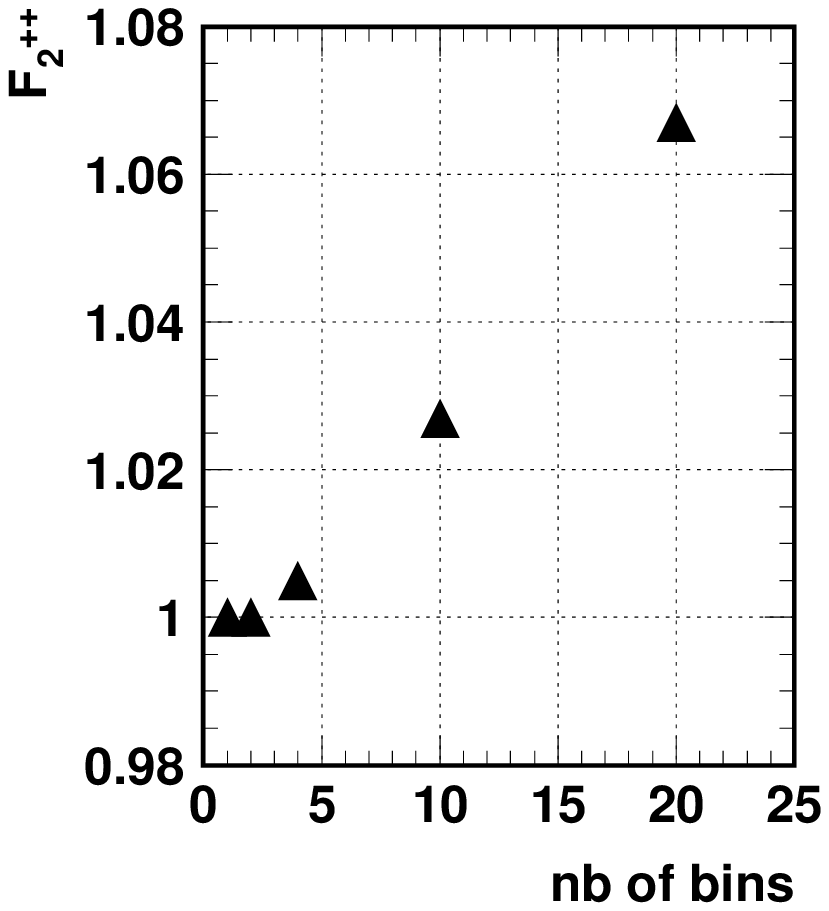}
    \includegraphics[width=0.4\textwidth]{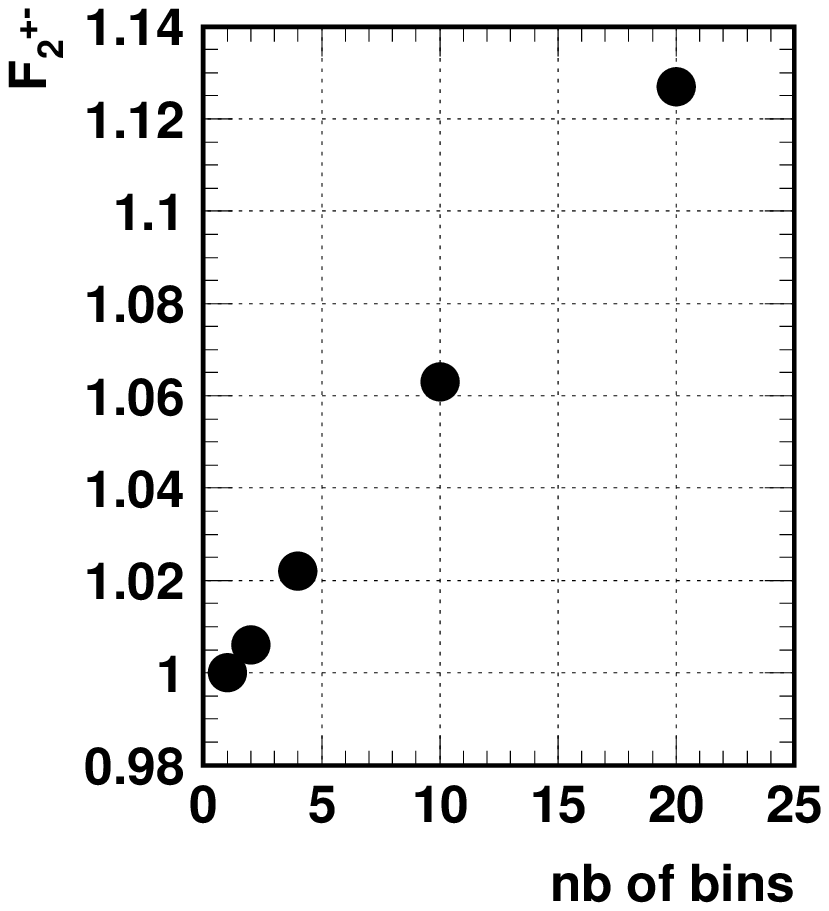}
  \end{center}
  \caption{Euro future contract. Like-sign and unlike-sign factorial moments $F_2^{++}$ 
and $F_2^{+-}$ are displayed for $1$, $2$, $4$, $10$ or $20$ bins.
The global time window of $16.5$ hours ($n_{bins}=1$).
This provides a time resolution ranging from $1.6$ hours for $20$ bins till $16.5$ hours for $1$ bin.
We observe that for $1$ and $2$ bins segmentation, $F_2^{++}$ is found equal to $1$
and $F_2^{++}$ is increasing above $1$ when the number of bins gets larger than $4$.
This confirms that non-Gaussian fluctuations in the sequence of returns
returns start to play a role when the resolution
is below $4$ hours. }
\label{f2pp}
\end{figure}

As explained in previous sections, correlations
between returns may lead to a broadening of the multiplicity distributions 
($n_+$ or $n_-$ or even a combination of both) 
and to dynamical fluctuations. In this case, the  factorial 
moments may increase  
with decreasing $\Delta$, or increasing the number of bins that divide $\Delta$. 
In Ref. \cite{lolo}, we have used the factorial moment of
second order $F_2$
for like-sign returns
\be
F_{2}^{++}=\frac{1}{N_{events}}\sum_{events}
\frac{\sum_{k=1}^{n_{bins}} \left\{n_{k}^{+}(n_{k}^{+}-1)
\right\}/n_{bins}}
{(\langle n^{+}\rangle /n_{bins})^{2}}
\label{like}
\ee
where $\langle n \rangle$ is the average number of positive returns in the full
time window  ($\Delta$), $n_{bins}$ denotes the number of bins in
this window and $n_k^{+}$ is the number of positive returns in $k$-th bin.
Similarly, we have defined the unlike-sign returns for $F_2$
\be
F_{2}^{+-}=\frac{1}{N_{events}}\sum_{events}
\frac{\sum_{k=1}^{n_{bins}} \left\{n_{k}^{+}n_{k}^{-}
\right\}/n_{bins}}
{\langle n^{+}\rangle \langle n^{-}\rangle /(n_{bins})^{2}}
\label{unlike}
\ee
with similar notations.
Considering different price series, we have shown that
for a small resolution in time window \cite{lolo}, below $4$ to $8$ hours, 
a deviation with respect to pure Gaussian fluctuations is observed from
the shape of second order factorial moments (\ref{like}) and (\ref{unlike})
\cite{lolo}.
Results are reminded in Fig. \ref{f2pp}.

\begin{figure}[htbp]
 \begin{center}
   \includegraphics[width=0.7\textwidth]{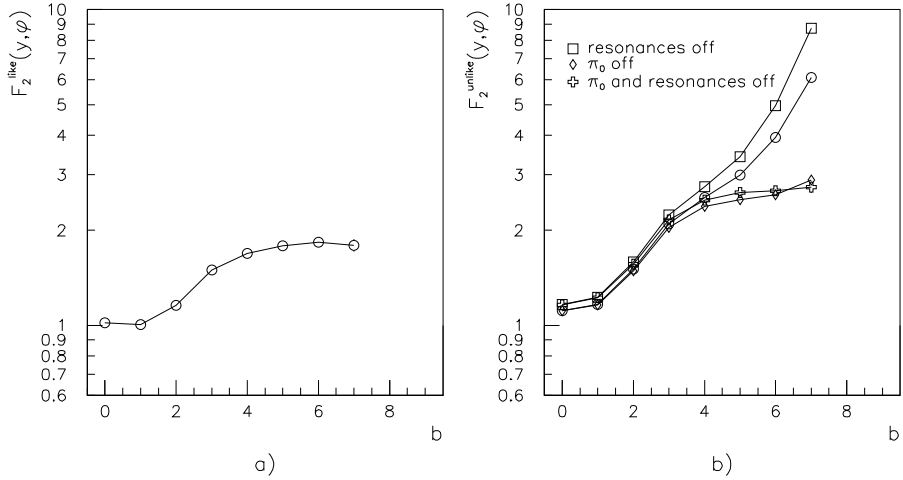}
 \end{center}
\caption{$e^+e^-$ collisions at high energy. Factorial moments $F_2(y,\varphi)$
 for like-charge (a) and unlike-charge (b)
 factorial moments $F_2^{++}$ 
and $F_2^{+-}$, calculated using JETSET 7.4 for a center of mass energy of $91.5$ GeV.
  See text for details.}
\label{rames}
\end{figure}

In order to compare the strength of the intermittency obtained in
financial data, it is interesting to provide a comparison with
standard high energy phenomenon in particle physics \cite{d1,rames}.
In what follows we discuss the results obtained with 
two of the most widely used $e^+e^-$ MC event generators, JETSET
\cite{Jetset} and HERWIG \cite{Herwig}. As it is done in the finance case,
we define moments for like-charge and
unlike-charge combinations of particles separately.
Fig. \ref{rames} shows the behavior of  $F_{2}^{++,+-}(n_{bins})$
at a centre-of-mass energy $91.5$ GeV \cite{rames}.
Without entering into details, we observe the rise of 
both observables with $n_{bins} \equiv b$ as in Fig. \ref{f2pp}.

By comparing Fig. \ref{f2pp} and Fig. \ref{rames}, we notice also
that the intermittency is weak in the financial data w.r.t. high energy data,
which is a reasonable observation.

\section{Gap Probability}

\begin{figure}[htbp]
 \begin{center}
   \includegraphics[width=0.45\textwidth]{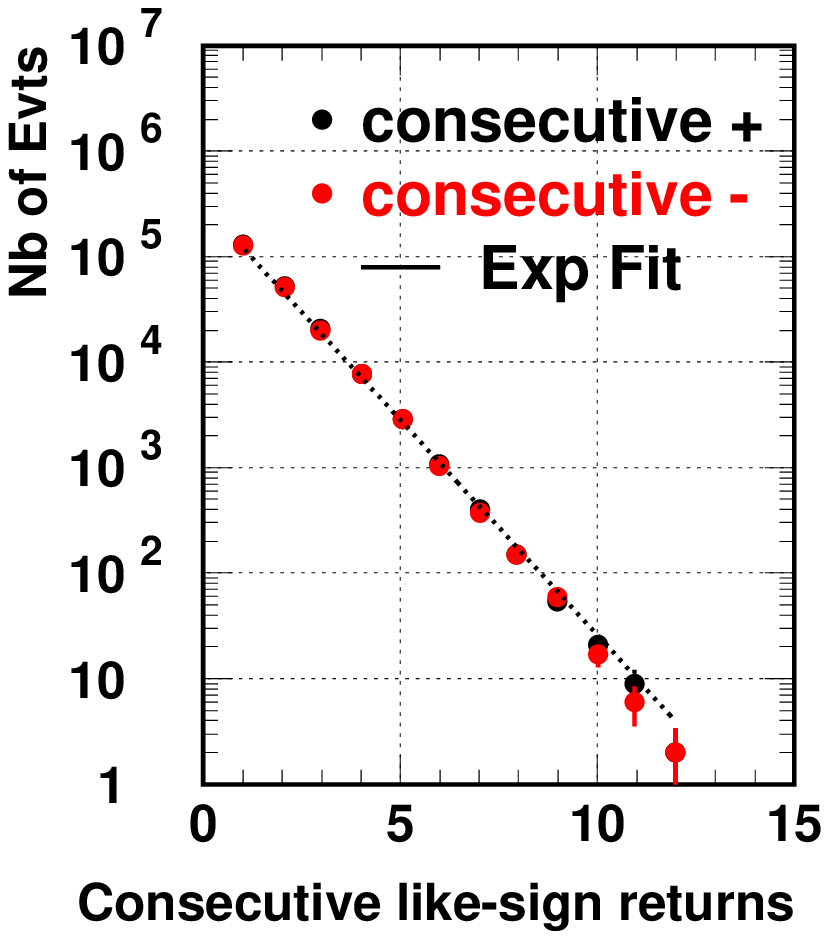}
   \includegraphics[width=0.45\textwidth]{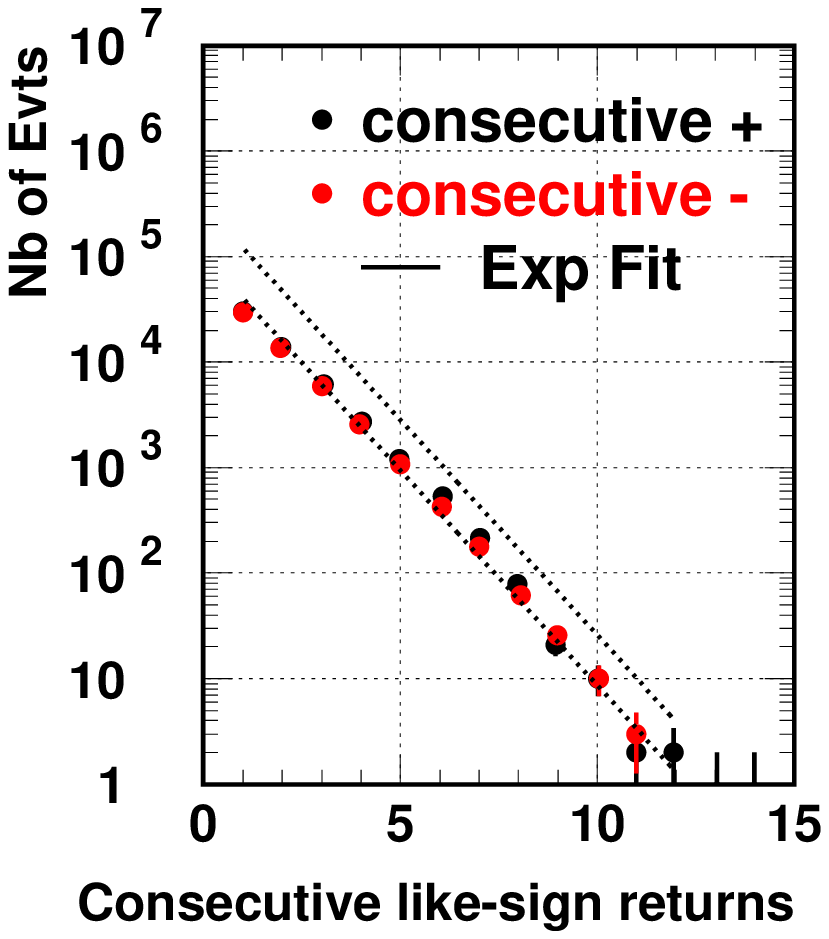}
 \end{center}
\caption{Left: Euro future contract sampled in 5min time units.
Right: Euro future contract sampled in 30min time units.
The distribution of events (probability distribution) is displayed as a function of the
size of the gap.
The gap is defined as the number of consecutive positive returns after a negative return,
thus this is a gap in negative return. Inversely for a gap in positive return.
The Euro future contract
 series over 10 years is sampled in two different time units (5min-Left and 30min-Right).
In both cases, we observe effectively an exponential fall of the probability distribution
as a function of the gap size. This confirms the relation derived in this letter
$
p_0 \simeq \exp(-K_1) = \exp(-<n_{+,-}>) = \exp(-\rho \Delta)
$.
We have presented an exponential fit on top of each plot. On the Right plot (30 minutes sampling),
we super-impose also the fit obtained on the 5 minutes sampling, in order to show that
the exponential slope is identical in both cases.
}
\label{gap}
\end{figure}

We can use the formalism introduced in section \ref{formalism}
in order to derive some further statements. From Eq. (\ref{7}),
we get
$$
G(-1)=p_0
$$
which corresponds to the probability to have zero particle in a phase space $\Delta$
or to have zero positive (resp. negative) returns in a given time window (finance case).
This defines a gap, either in rapidity for particles or in duration for 
positive (resp. negative) returns.
Let us use Eq. (\ref{5}) to express $G(-1)$ in another way
using cumulants $K_q$
$$
G(-1)=\exp(-K_1 +K_2/2! +...)
$$
When we can neglect correlations within a large time window \cite{lolo},
we have $K_2=0$ and we conclude
\be
p_0 \simeq \exp(-K_1) = \exp(-<n_{+,-}>) = \exp(-\rho \Delta)
\label{final}
\ee
where $\Delta$ represents either the phase space or the time window. 
This last expression is very simple and instructive. It states that the  gap
probability is exponentially suppressed either in rapidity or in time.
This is illustrated in Fig. \ref{gap} for the finance case.
We observe the distribution of events (probability distribution) as a function of the
size of the gap. 
The gap is defined as the number of consecutive positive returns after a negative return,
thus this is a gap in negative return. Inversely for a gap in positive return.
This gap is given in number of time units for the financial time series
considered. In Fig. \ref{gap}, we display results for  the Euro future contract
 series over 10 years, sampled in two different time units.
In both cases we observe effectively an exponential fall of the probability distribution
as a function of the gap size.  As illustrated also in Fig. \ref{gap},
this exponential fall does not depend on the sampling.
This confirms the prediction of Eq. (\ref{final}). The case of particle physics is
more complex, see Ref. \cite{lolo2}.

\section{Conclusion}

In this letter, we have discussed a direct comparison between high energy
physics  and quantitative finance results on factorial moments analysis.
Both for physics and finance, we have illustrated that correlations
between particles lead to a broadening of the multiplicity distribution 
and to dynamical fluctuations when the resolution becomes small enough.
From the generating function of factorial moments, we 
have shown that $p_0 \simeq  \exp(-\rho \Delta)$. 
This expression states that the gap
probability is exponentially suppressed within the gap size.
The gap is defined as the number of consecutive positive returns after a negative return,
thus this is a gap in negative return. Inversely for a gap in positive return.
We have confirmed this prediction with data.


\newpage

\end{document}